\documentclass[galaxies,article,accept,moreauthors,pdftex,10pt,a4paper,usenatbib]{mdpi}

\usepackage{booktabs}
\usepackage{multirow}
\usepackage{soul}
\usepackage{microtype}
\usepackage{upgreek}

\setitemize{parsep=6pt,itemsep=0pt,leftmargin=*,labelsep=5.5mm}
\setenumerate{parsep=6pt,itemsep=0pt,leftmargin=*,labelsep=5.5mm}
\setlist[description]{itemsep=0mm}

\firstpage{1}
\articlenumber{x}
\doinum{10.3390/------}
\pubvolume{xx}
\pubyear{2017}
\copyrightyear{2017}
\externaleditor{Academic Editors: Duncan A. Forbes and Ericson D. Lopeze}
\history{Received: 6 July 2017; Accepted: 11 August 2017; Published: date}

\usepackage{epsfig}
\usepackage{amsmath}
\usepackage{amssymb}
\usepackage{natbib}
\usepackage{multirow}
\usepackage{graphicx} 
\usepackage{color}
\usepackage{gensymb}
\usepackage{epstopdf}
\usepackage{mathtools}

\newcommand{\nat}{Nature} 

\newcommand{\eprint}[2][]{{\tt\if!#1!#2\else#1:#2\fi}}
\DeclarePairedDelimiter{\evdel}{\langle}{\rangle}
\newcommand{\ev}{\evdel}

\Title{On the Kinematics, Stability and Lifetime of Kinematically Distinct Cores: A Case Study}
\Author{{Felix Schulze} $^{1,{2},}$*, Rhea-Silvia Remus $^{1}$ and Klaus Dolag $^{1,{3}}$}

\AuthorNames{Felix Schulze, Rhea-Silvia Remus and Klaus Dolag}

\address{%
$^{1}$ \quad {Universit\"ats-Sternwarte M\"unchen}, Faculty of Physics, Ludwig-Maximilians-Universit\"at, Scheinerstr. 1, D-81679 M\"unchen, Germany; rhea@usm.uni-muenchen.de (R.-S.R.); dolag@usm.uni-muenchen.de (K.D.)\\

$^{2}$ \quad Max Planck Institute for Extraterrestrial Physics, Giessenbachstrasse 1, D-85740 Garching, Germany\\
$^{3}$ \quad Max Planck Institute for Astrophysics, Karl-Schwarzschild-Str. 1, D-85748 Garching, Germany
}
\corres{Correspondence: fschulze@usm.lmu.de}

\abstract{We present a case study of a early-type galaxy  {(ETG)} hosting a kinematically distinct core (KDC) formed in a binary high resolution 1:1 spiral galaxy merger simulation. The runtime of the simulation is pushed up to $10~\mathrm{Gyr}$ to follow the complete evolution of various physical properties. To investigate the origin of the KDC, the stellar component residing within the KDC is dissected, revealing that the rotational signal is purely generated by stars
that belong to the KDC for at least $0.5~\mathrm{Gyr}$ and are newly formed during the merging process. Following the orientation of the total
stellar angular momentum of the KDC, we show that it performs a motion comparable to the precession of a gyroscope in a gravitational potential.
We draw the conclusion that the motion of the KDC is a superposition of an intrinsic rotation and a global precession that gets gradually damped over cosmic time.
Finally, the stability of the KDC over the complete runtime of the simulation is investigated by tracing the evolution of the widely used $\lambda_{R}$ parameter and the misalignment angle distribution. We find that the KDC is stable for about $3~\mathrm{Gyr}$ after the merger and subsequently disperses completely on a timescale of $\approx$1.5~$\mathrm{Gyr}$.
}


\begin{document}

\section{Introduction}
The stellar kinematics of galaxies represent a meaningful benchmark for modern models of galaxy formation. Recent advances in
integral field spectroscopy revealed a rich variety of kinematical features in the line-of-sight velocity distribution (LOSVD), especially
in {early-type galaxies} (\mbox{Emsellem et~al.,~2004~\cite{2004MNRAS.352..721E}};
Krajnovi{\'c} et~al.,~2011~\cite{2011MNRAS.414.2923K}). These features embody the final state of a complex assembly history and evolution shaping the dynamical and kinematical appearance of galaxies.
A~particularly interesting class of kinematic appearances are galaxies that exhibit kinematically distinct cores (KDCs), which are kinematically decoupled from their host galaxy, often visible in an inclined net rotation of the central core component. Providing full two-dimensional observations of the stellar kinematics of statistically meaningful samples, the $\mathrm{ATLAS}^{3D}$ (Cappellari et~al.,~2011~\cite{2011MNRAS.413..813C}) and CALIFA (S{\'a}nchez et~al.,~2012~\cite{2012A&A...538A...8S}) surveys unveiled a significant fraction of {ETGs}
exhibiting KDCs.

Results from McDermid et~al.,~2006 \cite{2006NewAR..49..521M}, who investigated the central region of {ETGs} using the OASIS spectrograph, suggest two fundamental types of KDCs: the first type are KDCs that exhibit an old stellar population (>$8~\mathrm{Gyr}$) contemporary to the surrounding host galaxy. This indicates that those KDCs are not a result of recent merging and were more likely to be formed through accreted material or merging at earlier times. KDCs of this type are typically {extended to} $\mathrm{kpc}$ scale while residing in non-rotating galaxies. The
second type of KDCs are comprised of a more compact younger stellar population extending characteristically out to a few $100~\mathrm{pc}$. These KDCs characteristically exist in fast rotating galaxies emphasising the different formation histories of the two types.

A diagnostically conclusive method to probe the general formation pathways of KDCs is to utilise binary merger simulations of spiral galaxies. Within this
framework, it is possible to follow the evolution of galaxy properties in great detail. Early theoretical studies suggest that young KDCs can arise from  {a} binary galaxy merger on a retrograde orbit via in situ star formation ({Balcells et~al.,~1990}~\cite{1990ApJ...361..381B};
Hernquist et~al.,~1991~\cite{1991Natur.354..210H}). In a more recent
study, Tsatsi et~al.,~2015~\cite{2015ApJ...802L...3T} showed that old KDCs can also originate from a initially prograde merger through a reversal of the orbital spin induced by reactive forces due to substantial mass loss. Furthermore,  {Hoffman et~al.,~2010~\cite{2010ApJ...723..818H}} showed that the initial gas fraction ($f_{gas}$) of the progenitors has  {a} substantial impact on the existence of a  {KDC} in the  {center of the} merger remnant. Analysing a sample of $56$ 1:1 binary spiral mergers with varying orbital parameters and $f_{gas}$, they show that, for $f_{gas} < 10 \%$, the remnants do not host a KDC, while, for $10 \% < f_{gas} < 40 \%$, the
fraction of remnants hosting a KDC increases. However, while the origin of young KDCs seem to be well understood by now, their stability and lifetime has not been studied in great detail.

In this work we present a case-study for the evolution of a second-type KDC formed in a major merger event, demonstrating that these structures might only
be visible for a short timespan.
\section{Simulation}
We perform a case study of a single binary merger simulation selected from a sample  {of $10$ high resolution simulations} outlined in Schauer et~al.,~2014~\cite{2014ApJ...783L..32S}.
 {From the $10$ simulations, only two have a KDC: none of the 3:1 mergers show any sign of a KDC, while the presence of a bulge does not seem to
influence the existence of a KDC. In fact, the two simulations that exhibit a KDC are identical in all configurations except for the inclusion of a bulge in the
progenitor spiral galaxies. Additionally, albeit our sample is small, we also see a coupling between the initial gas fraction $f_{gas}$ and the appearance
of a KDC, similar to the results presented by Hoffman et~al.,~2010~\cite{2010ApJ...723..818H}. From the different orbital configurations in this sample,
only one leads to a KDC. This suggests that the orbital configuration of the initial merger setup is important for the formation of a KDC; however, this is
not studied in this work.}

Using the TreeSPH-code GADGET2 (Springel et~al.,~2005~\cite{2005MNRAS.364.1105S}), all simulations implement various physical processes like
star formation, supernova feedback (Springel et~al.,~2003~\cite{2003MNRAS.339..289S}) and black hole feedback (Springel et~al.,~2005~\cite{2005MNRAS.361..776S}).
We use the 1:1 spiral-spiral merger that manifests the following orbital configuration: inclinations $i_1=-109 \degree$ and $i_2=180  \degree$, pericenter
arguments $\omega_1=60  \degree$ and $\omega_2=0  \degree$ according to \cite{1972ApJ...178..623T} (for further information about the simulation, see
Johansson et~al.,~2009b~\cite{2009ApJ...690..802J}, Johansson et~al.,~2009a~\cite{2009ApJ...707L.184J} and Remus et~al.,~2013~\cite{2013ApJ...766...71R}). The two spiral progenitors are identical clones with an initial gas fraction of $20 \%$ hosting a stellar bulge as well as a central black hole. In order to investigate the stability and kinematics of the KDC, we trace several properties like LOSVD, stellar and gaseous angular momentum, mass and star formation rates (SFR) of the KDC in time as well as those of the hosting galaxy. The simulation is run up to $10.2~\mathrm{Gyr}$.
The simulation allows to subdivide the total stellar population into the two subpopulations of initial stars already present in the progenitor galaxies, and stars formed during the simulation runtime.
The gravitational softening length for stars formed from the gas during the simulation is $\epsilon=0.1~\mathrm{kpc}/$h, while the softening for the initial stellar component is set to $\epsilon=0.2~\mathrm{kpc}/$h.  {This choice ensures that the maximum gravitational force exerted from a particle does not depend on its mass (Johansson et~al.,~2009b~\cite{2009ApJ...690..802J}).}

The selection of this particular merger for this study is based on a visual inspection of the LOSVD at a simulation time of $2.7~\mathrm{Gyr}$ (see Figure \ref{fig:kdc_map_snap_134}), revealing the presence of a counter rotating kinematically distinct core in the centre. Throughout this study, LOSVD maps are Voronoi-binned using the method outlined in Cappellari \& Copin, 2003~\cite{2003MNRAS.342..345C}, ensuring a minimum of $100$ particles per cell to reduce the statistical {uncertainty} of the mean velocity.

The binary merger proceeds as follows: subsequent to the approaching phase lasting until $t=0.66~\mathrm{Gyr}$, a rapid merging phase with two encounters follows. The first encounter takes place at $t_1=0.66~\mathrm{Gyr}$, while
the second and final encounter happens at $t_2=1.3~\mathrm{Gyr}$. Afterwards, the remnant relaxes under the influence of dynamical friction and violent relaxation.
\section{Results}\vspace{-6pt}

\subsection{Global Properties}
\begin{figure}[H]
\centering
\includegraphics[width=14 cm]{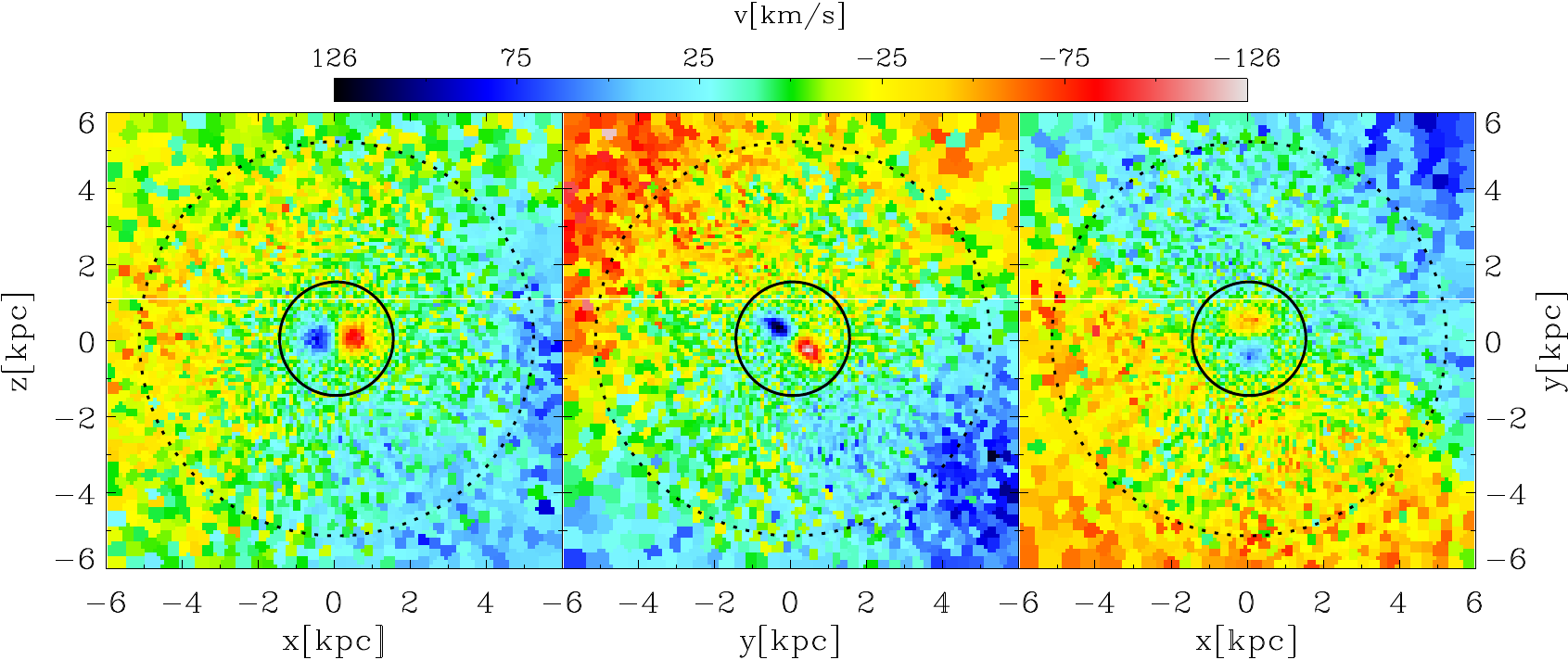}
\caption{Each panel displays {a} {LOSVD} map of the central region of the merger remnant
hosting the {KDC} in different projections  {at $t=2.7~\mathrm{Gyr}$}. From left to right:
$x$--$z$ plane, $y$--$z$ plane and $x$--$y$ plane. The color bar on the top indicates the velocity scaling adapted individually for each panel. The black solid circle marks the estimated KDC radius of $1.5~\mathrm{kpc}$, while the dashed circle indicates the half-mass~radius.}
{\label{fig:kdc_map_snap_134}}
\end{figure}

Before investigating the kinematics of the KDC in detail, we have to determine the spatial extent of the KDC. Figure \ref{fig:kdc_map_snap_134} shows the LOSVD of the merger remnant projection onto the three coordinate planes after a simulation runtime of $2.7~\mathrm{Gyr}$ that is $1.5~\mathrm{Gyr}$ after the second encounter.
As can clearly be seen, the merger remnant hosts a distinct central rotating feature that is present in all three~projections (i.e., never seen face-on) and is clearly misaligned to the rotation of its surroundings. From~the visual appearance of the LOSVD map, we conservatively estimate the radius of the KDC to be $1.5~\mathrm{kpc}$, as~shown by the solid black circle {in Figure} \ref{fig:kdc_map_snap_134}.
{This corresponds} roughly to a third of the  {stellar} half-mass radius ($r_{1/2}=5.2~\mathrm{kpc}$) of the  {remnant}, which is illustrated by the black dashed line, and~therefore much larger than the resolution limit.  {The half-mass radius is determined to be the
radius of a three-dimensional sphere containing half of the total stellar mass centred on the galaxy center of~mass.}

\begin{figure}[H]
\centering
\includegraphics[width=11.5 cm]{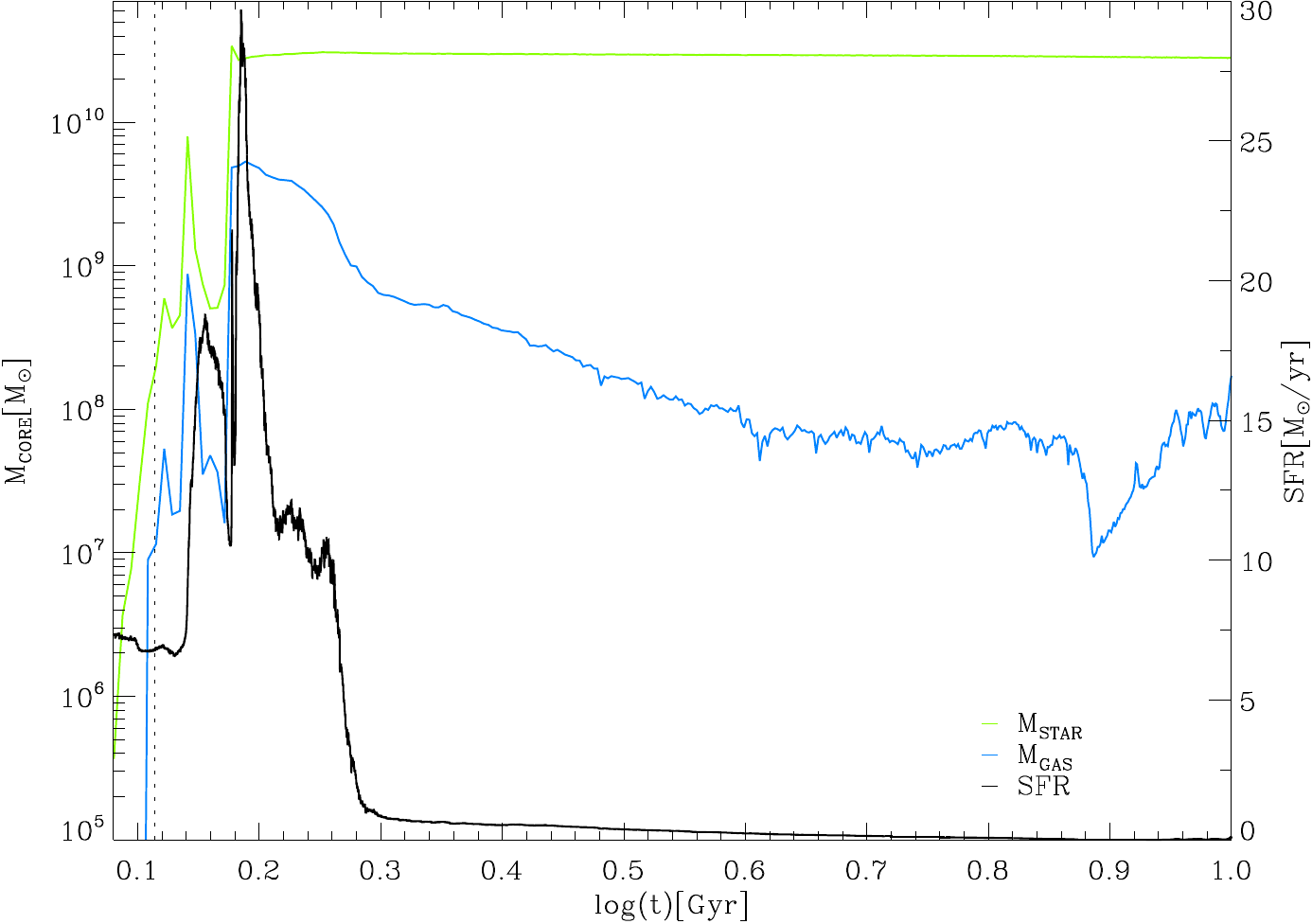}
\caption{Temporal evolution of global core properties and the total star formation rate (SFR) of the complete galaxy. The black line indicates the
{SFR} of the entire galaxy, while the green and blue curves represent the total stellar and gaseous mass within the KDC, respectively. The vertical dashed line marks the moment of the second encounter.}
{\label{fig:kdc_global}}
\end{figure}

Figure \ref{fig:kdc_global} displays the temporal evolution of the stellar (green) and gaseous (blue) mass within the core as well as the total
SFR of the galaxy in black. The starting point in time is chosen to be directly before the second encounter to also capture the starburst
triggered by the second encounter. This~starburst lasts for approximately $0.7~\mathrm{Gyr}$, with a maximum SFR {of }$28~\mathrm{M_{\odot}/yr}$.

During this period, the stellar mass inside the KDC increases until it plateaus at $ \approx 3 \cdot 10^{10}~\mathrm{M_{\odot}}$. Afterwards, $M_{STAR}$ stays constant, excluding significant infall of stars. Consistently with a starburst, the amount of gas in the KDC
decreases drastically by roughly one order of magnitude within this timeframe. The subsequent modest decrease of $M_{GAS}$ suggest ongoing
star formation activity within the KDC, however at much lower rates.

\subsection{Dissecting the KDC} \label{dissect}
To investigate the origin of the KDC signal in the LOSVD maps, we explore the contributions of stellar populations with different properties to the signal. We distinguish between stars that are set up in the initial conditions and stars that are produced during the runtime of the simulation, denoting them \textit{'Initial Stars (IS)'} and \textit{'Newly Formed Stars (NFS)'}. The initial gas from which the NFS are formed is expected to form a disc in the centre during the merger due to its hydrodynamical nature and therefore generates a orbital configuration characteristic for discs. In contrast, the IS are only  {affected} by violent relaxation and dynamical friction. Of course, the interplay between those processes and their efficiency is complex and depends on the parameters of the merger and hence cannot be predicted easily.

In addition, we split the stars inside the core into two groups,  {independently of whether they} are IS or NFS: stellar particles that are permanently located within the KDC are classified as \textit{'Permanent Core Stars (PCS)'}, and stars which are localised only temporarily within the core denoted \textit{'Temporary Core Stars (TCS)'}. TCS are expected to move on highly elliptical orbits with apocenters beyond the KDC radius.
For this purpose, we use a algorithm that iteratively removes particles that move outside of the KDC. At a runtime of $2.2~\mathrm{Gyr}$, where the
KDC is fully evolved, the algorithm selects all stellar particles within the KDC radius, traces this parent sample of stars to the consecutive
snapshot, and removes the particles that leave the KDC radius from the sample. This procedure is reiterated for $24$ subsequent  {snapshots} until a runtime
of $2.7~\mathrm{Gyr}$. In this manner, the algorithm creates a sample of stellar particles that reside within the KDC for at least $0.5~\mathrm{Gyr}$. This corresponds to approximately one characteristic orbital period within the KDC.
By construction, this algorithm also disregards stellar particles falling into the KDC during the application of the procedure from the sample.

\begin{figure}[H]
\centering
\includegraphics[width=14 cm]{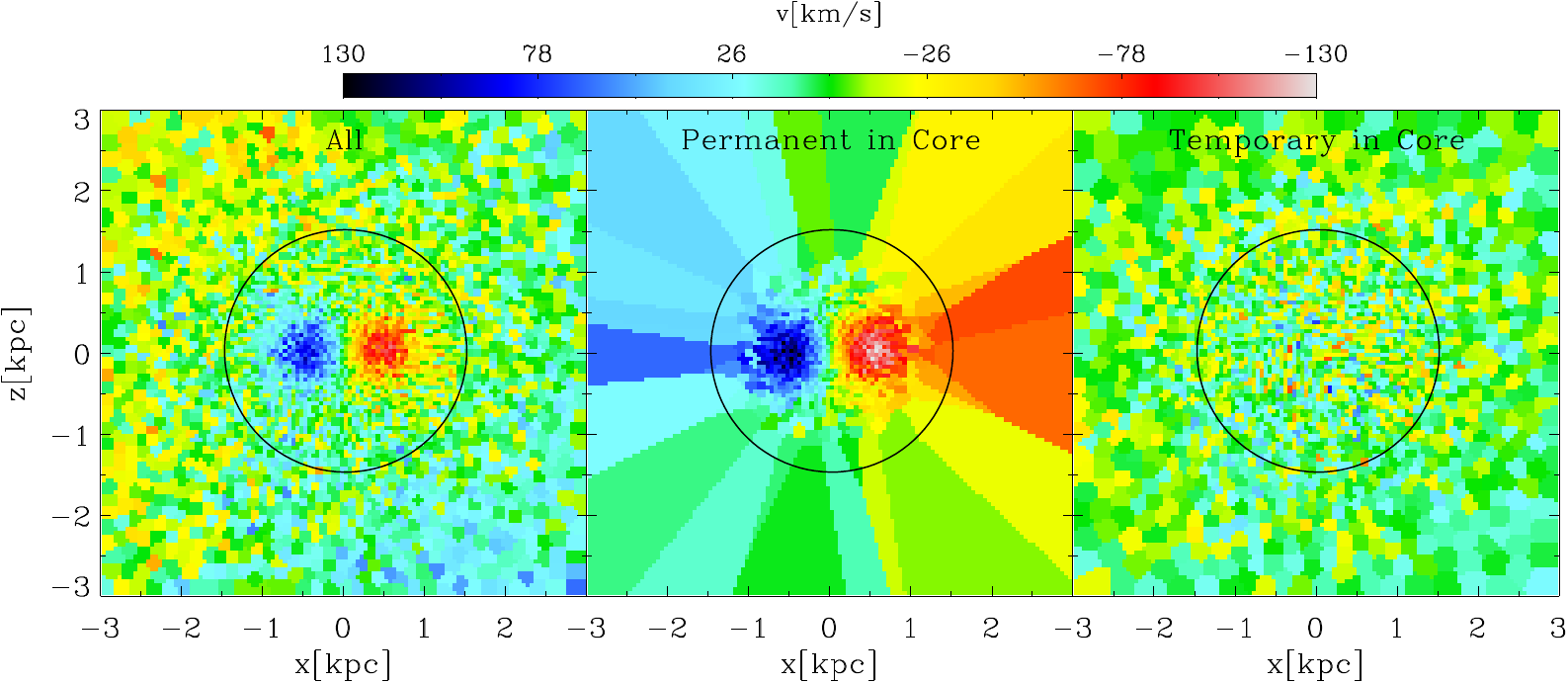}
\caption{\textbf{Left panel}: LOSVD of the central $3~\mathrm{kpc}$ in the $x$--$z$ projection considering all stars in the line-of sight; \textbf{central panel}: LOSVD map of the central $3~\mathrm{kpc}$ in the $x$--$z$ projection taking into account stars that are determined to belong permanently to the KDC; \textbf{right panel}: LOSVD map of the central $3~\mathrm{kpc}$ in the $x$--$z$ projection including stars that only temporarily reside in the core. The panels share a common velocity scaling given in the color bar. The solid black circle indicates the core radius of $r_{CORE}=1.5~\mathrm{kpc}$.}
{\label{fig:kdc_map_snap_134_dissect_1}}
\end{figure}

Figure \ref{fig:kdc_map_snap_134_dissect_1} directly compares the visual contribution of the PCS and TCS to the rotational signature in the LOSVD map
at $t=2.7~\mathrm{Gyr}$.

The left panel displays a zoom-in onto the KDC including all stellar particles in the line-of-sight. Velocity maps for the TCS and PCS are shown in the central and right panel, respectively.
As can clearly be seen, the rotational signal is the net rotation induced by the PCS component overlaying the TCS component, which is dominated by random motion. The fact that the maximum velocities reached by the PCS is higher than those of the full stellar sample infers that the random moving component diminishes the signal as expected.

\begin{figure}[H]
\centering
\includegraphics[width=14 cm]{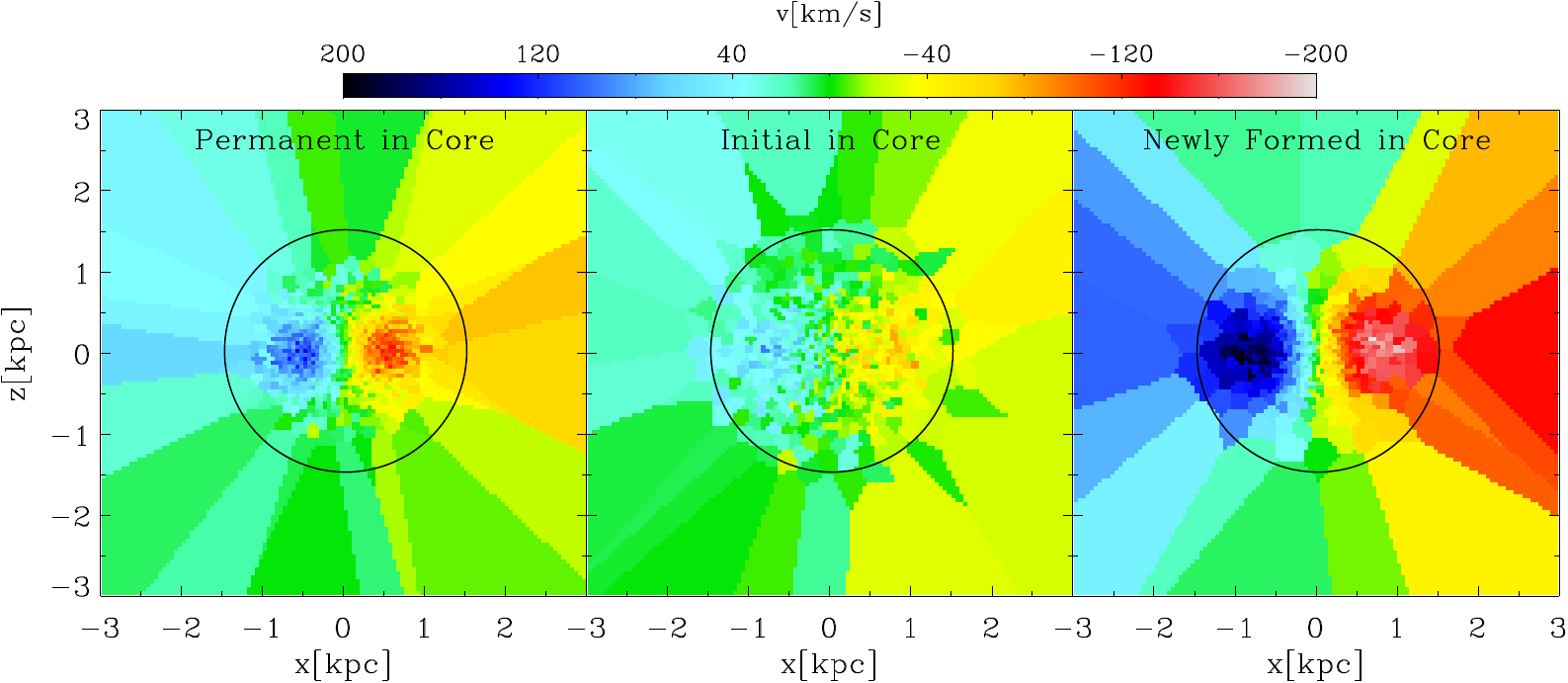}
\caption{LOSVD maps of the central $3~\mathrm{kpc}$ in the $x$--$z$ projection. \textbf{Left panel}: all permanent core stars; \textbf{central panel}: permanent core stars that are initially in one of the progenitors; \textbf{right panel}: permanent core stars that are formed in situ during the simulation. The panels share a common velocity scaling given in the color bar.
The solid black circle indicates the core radius of $r_{CORE}=1.5~\mathrm{kpc}$.}
{\label{fig:kdc_map_snap_134_dissect_2}}
\end{figure}

As we are interested in the origin of the KDC, we further separate the PCS into the two populations of IS and NFS. The result is displayed in
Figure \ref{fig:kdc_map_snap_134_dissect_2}: the left velocity map replicates the central panel of Figure \ref{fig:kdc_map_snap_134_dissect_1}, while
the central and right panels show the subdivision into IS and NFS, respectively. Comparing the central and right panels demonstrates that the KDC is
dominated by the newly formed stars in agreement with the results from Hoffman et~al., ~2010~\cite{2010ApJ...723..818H}.
The NFS component exhibits a peak velocity that is a roughly a factor of two higher than the peak velocity of the IS component.
We~reason that the minor rotation in the IS component might be generated by the drag caused by the fast rotating~particles.

From our analysis we can conclude that the KDC signal in the merger remnant is generated mainly by stellar particles that permanently reside within the
core and are formed during the merging process in situ. Therefore, the KDC seems to be in agreement with the younger more compact KDCs found observationally by McDermid et~al., 2006~\cite{2006NewAR..49..521M}.


\subsection{Evolution of Stellar KDC Kinematics} \label{kinematics}
\begin{figure}[H]
\centering
\includegraphics[width=13.0 cm]{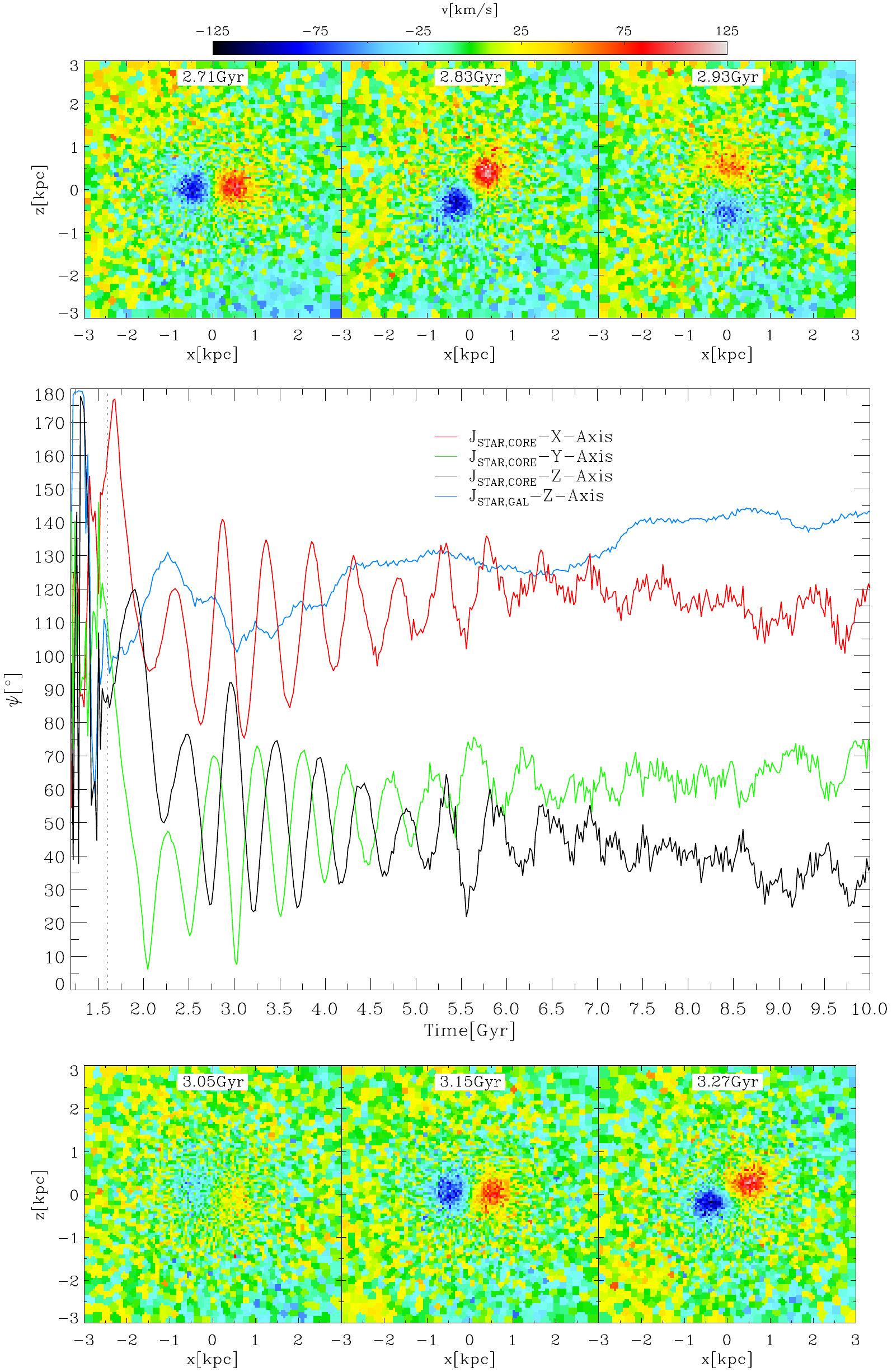}
\caption{The main panel shows the temporal evolution of the angle between the total stellar angular momentum of the KDC and the three coordinate axes as given in the legend. The blue line for comparison shows the evolution of the angle between the angular momentum of the surrounding galaxy and the $z$-axis. The dashed vertical line marks the time at which the stellar mass accretion of the KDC is finished (see Figure \ref{fig:oscillation_comp}). The upper and lower rows display a temporal sequence of the LOSVD map of the central $3~\mathrm{kpc}$ in the $x$--$z$ projection.}
{\label{fig:oscillation_comp}}
\end{figure}

To understand the evolution of the KDC, we investigate its kinematical behaviour over time with respect to the kinematical behaviour of the surrounding
host galaxy. The previous section revealed that the actual KDC is composed of stellar particles formed in situ
during the merger and hence might retain a fraction of the orbital angular momentum.
To measure the orientation of the KDC, we calculate the total angular momentum of all stars located inside the KDC ($J_{STAR,CORE}$) for each snapshot, and~calculate its angle to the three coordinate axes. In addition, the same angles are determined for the host galaxy to exclude a global mutual motion of the KDC and the
galaxy within the simulated box. The temporal evolution of the angles is shown in the main panel of Figure \ref{fig:oscillation_comp}. The~red, black and green curves represent the angles between the KDC and the coordinate axes, while the orange line marks the angle between the angular momentum
of the host galaxy and the $z$-axis.

The KDC angular momentum follows a completely unexpected behaviour: subsequent to a violent phase of relaxation lasting until $t=1.7~\mathrm{Gyr}$, the angles with respect to the coordinate axes
reveal an oscillation of the KDC, which gets gradually dampened in the further evolution.
The imprint of the oscillation is visible up to $t\approx 7~\mathrm{Gyr}$, and the period of the oscillation is nearly equal for each axis and is determined to be $\delta t\approx0.5~\mathrm{Gyr}$. As a two-dimensional visualisation, the upper and lower rows of Figure \ref{fig:oscillation_comp} show the velocity field in the $x$--$z$ plane projection at six points in times equally distributed over one oscillation period: from an initial alignment with the $x$-axis at $2.7 \mathrm{Gyr}$, the KDC turns upwards by $90 \degree$ at $t \approx 2.93~\mathrm{Gyr}$.
In the following snapshot, it gets more diffuse and almost disappears until it reverts to its initial configuration at $t=3.27~\mathrm{Gyr}$. The last panel shows the initiation of a new oscillation cycle indicated by the slight upturn.

\begin{figure}[H]
\centering
\includegraphics[width=15 cm]{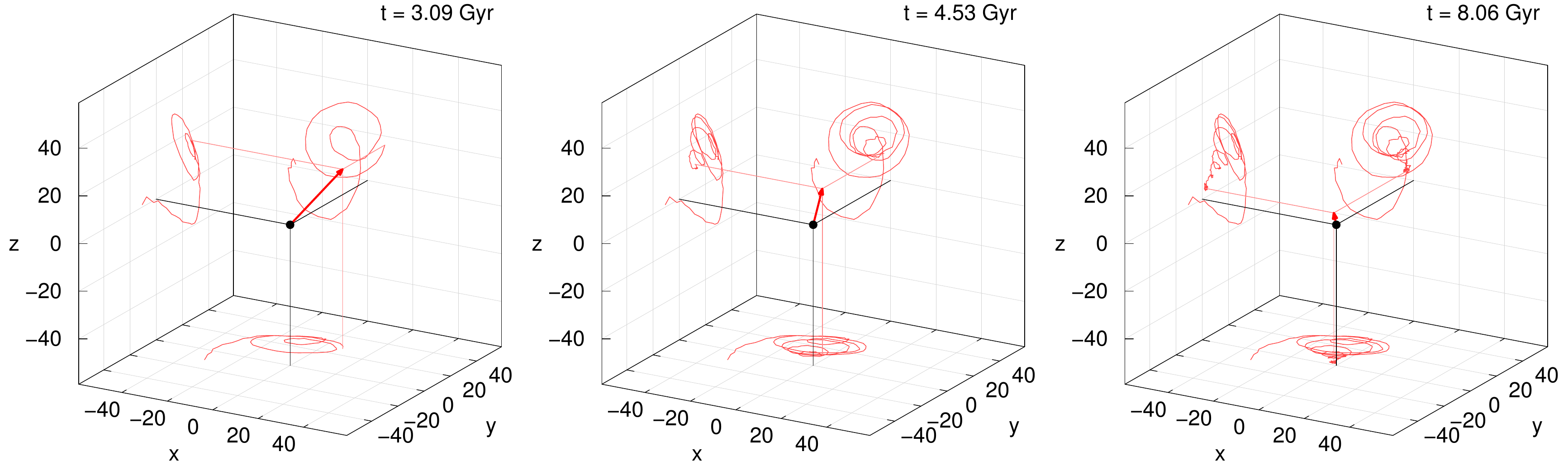}
\caption{Illustration of the direction of the total stellar angular momentum in three-dimensional space as a red vector at three different time steps, of the
simulation as given in the label. The red curves drawn on the coordinate planes trace the projected paths of the vector's endpoint until its current position.}
{\label{fig:movie_comp}}
\end{figure}

Of course, the projected appearance in the velocity maps is strongly influenced by the rotation of the KDC around the other two coordinate
axes. Furthermore, it is difficult to deduce the actual movement in three dimensions from the angles and the velocity maps. Therefore, Figure \ref{fig:movie_comp} illustrates the direction of the angular momentum vector of the KDC in three-dimensional space at three points in time. The curves drawn on the coordinate planes visualise the projected track of the vector's until its current position. In all three projections,
the tracks describe slightly distorted circles that get smaller due to the damping of the oscillation until the angular momentum stabilises in
direction.

From our analysis, we conclude that the total stellar angular momentum of the KDC performs a three-dimensional motion
comparable to the precession of a gyroscope in a gravitational potential superimposed with a general tilt of the precession axis.
From this, we infer that the kinematics of the KDC can be subdivided into an intrinsic rotation and a superimposed global figure rotation.

This result raises the questions of which mechanism generates this periodic, well-defined motion of the KDC and how it is damped. In the previous section we showed that the rotating component of the KDC  {comprises} stars formed during the merger and that
the intrinsic rotation  {is} the result of the process, whereby gas condenses into a disc.
We speculate that the precession is a residue of the orbital angular momentum of the merger event retained by the gaseous component.

%

\subsection{Fading of the KDC}
After revealing the origin and kinematics of the KDC in the previous sections, we now examine the temporal stability of the KDC with regard to its kinematical properties. The timescale on which the young compact KDCs are stable is still widely unknown since only numerical simulations with sufficient resolution can trace their evolution. In contrast, classical old KDCs are expected to be stable for up to $10~\mathrm{Gyr}$ since their stellar population is indistinguishable from their host galaxy.

One of the few investigations of KDC stability was conducted by McDermid et~al.,~2006~\cite{2006MNRAS.373..906M} using a simulation based on {SSP} model spectra. They find
that the KDC is fading due to the stellar evolution of the young KDC stars, as their
luminosity-weighted contribution is diminished due to a increasing mass-to-light ratio. Following this argument, the KDC is still present, however forced to the background of the velocity field.

As shown in the previous section, our analysis provides a reasonable dissection of the core into its rotating and dispersion dominated component
by differentiating the IS and NFS components confirmed by their visual appearance in the LOSVD maps.

\begin{figure}[H]
\centering
\includegraphics[width=12.3 cm]{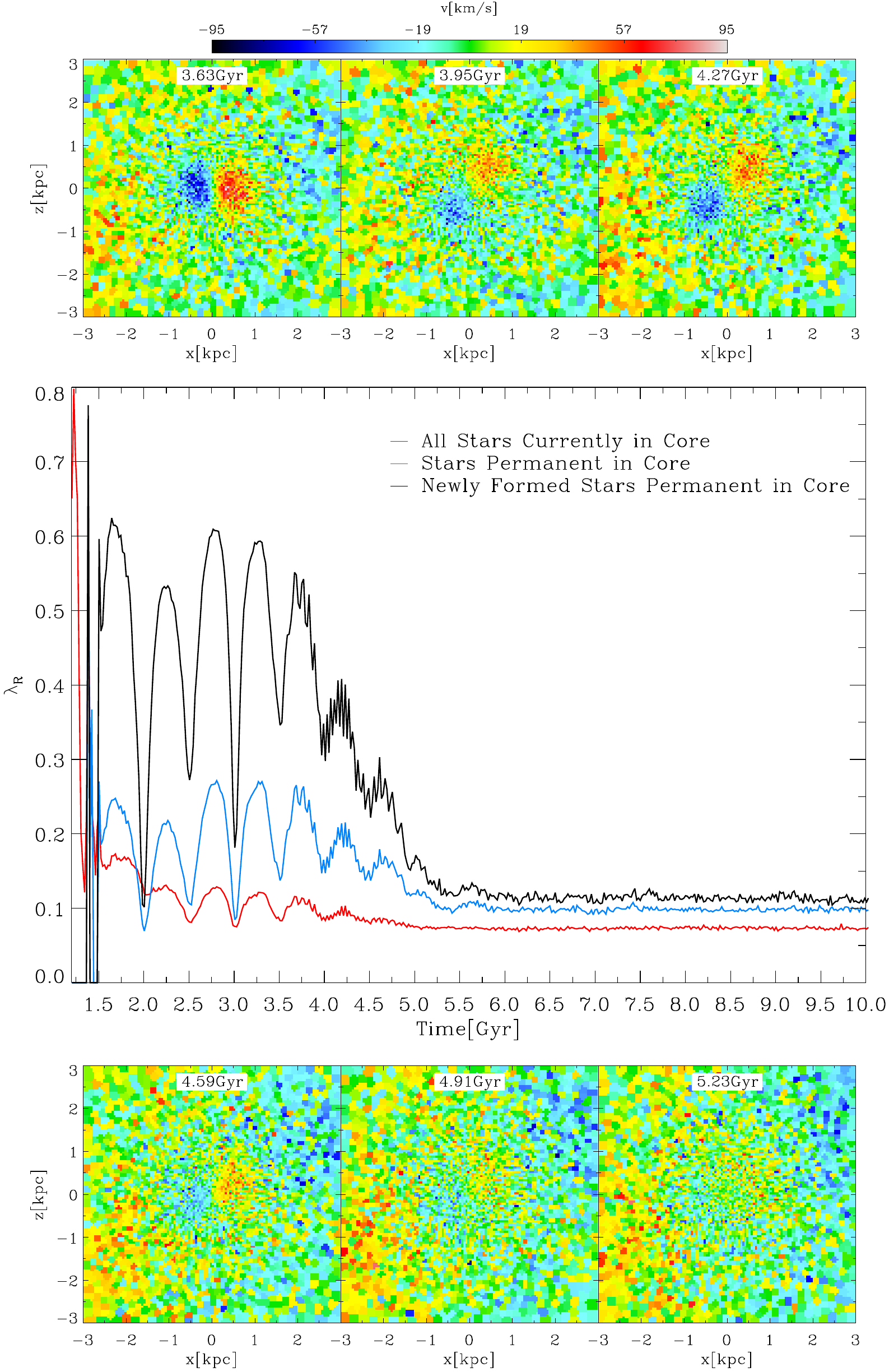}
\caption{The upper and lower rows display a temporal sequence of the LOSVD maps of the central $3~\mathrm{kpc}$ in the $x$--$z$ projection visualising the
gradual fade of the KDC. The main panel shows the temporal evolution of $\lambda_{R}$ in the $x$--$z$ projection split up into three stellar populations: All stars that reside inside the core at a given time are represented by the red curve, while the blue and black curve trace the evolution for the
permanent core stars and the newly formed permanent core stars, respectively.}
{\label{fig:l_r_core_map_comp}}
\end{figure}

A sequence of LOSVD maps in various decisive steps of the KDC evolution considering all stars in the line-of-sight from $t=3.6~\mathrm{Gyr}$ to $t=5.2~\mathrm{Gyr}$ are shown in the upper and lower rows of Figure~\ref{fig:l_r_core_map_comp}.
The~projection plane is held constant in all panels.
After a runtime of $3.63~\mathrm{Gyr}$, the KDC is still clearly visible performing a precession in the two following snapshots. Between $t=4.27~\mathrm{Gyr}$ and $t=5.33~\mathrm{Gyr}$, the KDC signal gradually gets
weaker until it vanishes almost completely. By testing different projections, we confirmed that this fading is real and not just due to a precession {of} the
core, as in the previous time steps where a weakening of the KDC signal could be detected in  {one} projection, while it strengthened in another projection.
Up to the full time of the simulation, the KDC does not build up again in any projection, clearly indicating that the KDC was dissolved.
Hence, we conclude that the KDC is fading on a timescale of $1.6~\mathrm{Gyr}$.

In order to quantify the rotational support of the stellar populations of the KDC, we calculate the $\lambda_{R}$ parameter for the core, which is given by
\begin{equation}
\lambda_{R}=\frac{\ev{R ~ \left| V  \right|}}{\ev{R ~ \sqrt{V^2+\sigma^2}}}=\frac{\sum_{i=1}^{N_{p}}F_{i} ~ R_{i} ~ |\overline V_{i}|}{\sum_{i=1}^{N_{p}}F_{i} ~ R_{i} ~ \sqrt{\overline V_{i}^2+\sigma_{i}^2}},
\end{equation}
where the summation runs over each pixel within the chosen aperture.  { $F_i$, $R_i$, $\left|\overline V_i \right|$ and $\sigma_i$ are the flux, projected distance to the galaxy centre, mean stellar velocity and velocity dispersion of {the} $i$th pixel, respectively (Emsellem et~al.,~2007~\cite{2007MNRAS.379..401E}). For simulations, the fluxes are replaced by stellar masses as we do not have luminosities, while assuming a constant mass-to-light ratio within each galaxy.} $\lambda_{R}$ is a measure of the ratio of random to ordered motion. Furthermore, $\lambda_{R}$  is an observationally accessible quantity, albeit sensitive to projection effects.

The temporal evolution of $\lambda_{R}$ in the core region, measured in the $x$--$z$ plane and considering different populations of stars, is displayed in the main panel of Figure \ref{fig:l_r_core_map_comp}.
The red curve shows $\lambda_{R}$ for all stellar particles within the KDC at each time step, while blue and black curves represent the PCS and newly formed PCS population, respectively. A common feature of all three populations is an imprinted oscillation, reflecting the projection effects induced by the
precession of the KDC revealed in the previous section.

Comparing the mean level of $\lambda_{R}$ for the three populations, the red and blue curves exhibit significantly less rotational support than the
black curve. For the total population, this is due to the included, highly dispersive TCS component. In the case of the
PCS, we showed that the IS subcomponent gets partly dragged into rotation, however, at a factor of two slower than the NFS subcomponent. Thus, the dispersion within the PCS is still rather large and explains the lower $\lambda_{R}$ values of the blue curve.

The newly formed PCS component shows $\lambda_{R}$-values as high as $\lambda_{R} \approx 0.6$, which is a clearly rotation-dominated signal.
However, due to the precession of the core, it can reach values as low as $\lambda_{R} \approx 0.1$ when the core precesses out of the $x$--$z$-plane,
extinguishing any rotational signal. Between $t=3.75~\mathrm{Gyr}$ and $t=5.25~\mathrm{Gyr}$, we find a gradual decrease to $\lambda_{R} \approx 0.13$,
which is clearly in the slow rotating regime populated by non-rotating early-type galaxies, on a timescale of $1.5~\mathrm{Gyr}$.
This is in agreement with the fading timescale inferred from the inspection of the LOSVD maps.

\begin{figure}[H]
\centering
\includegraphics[width=14.5 cm]{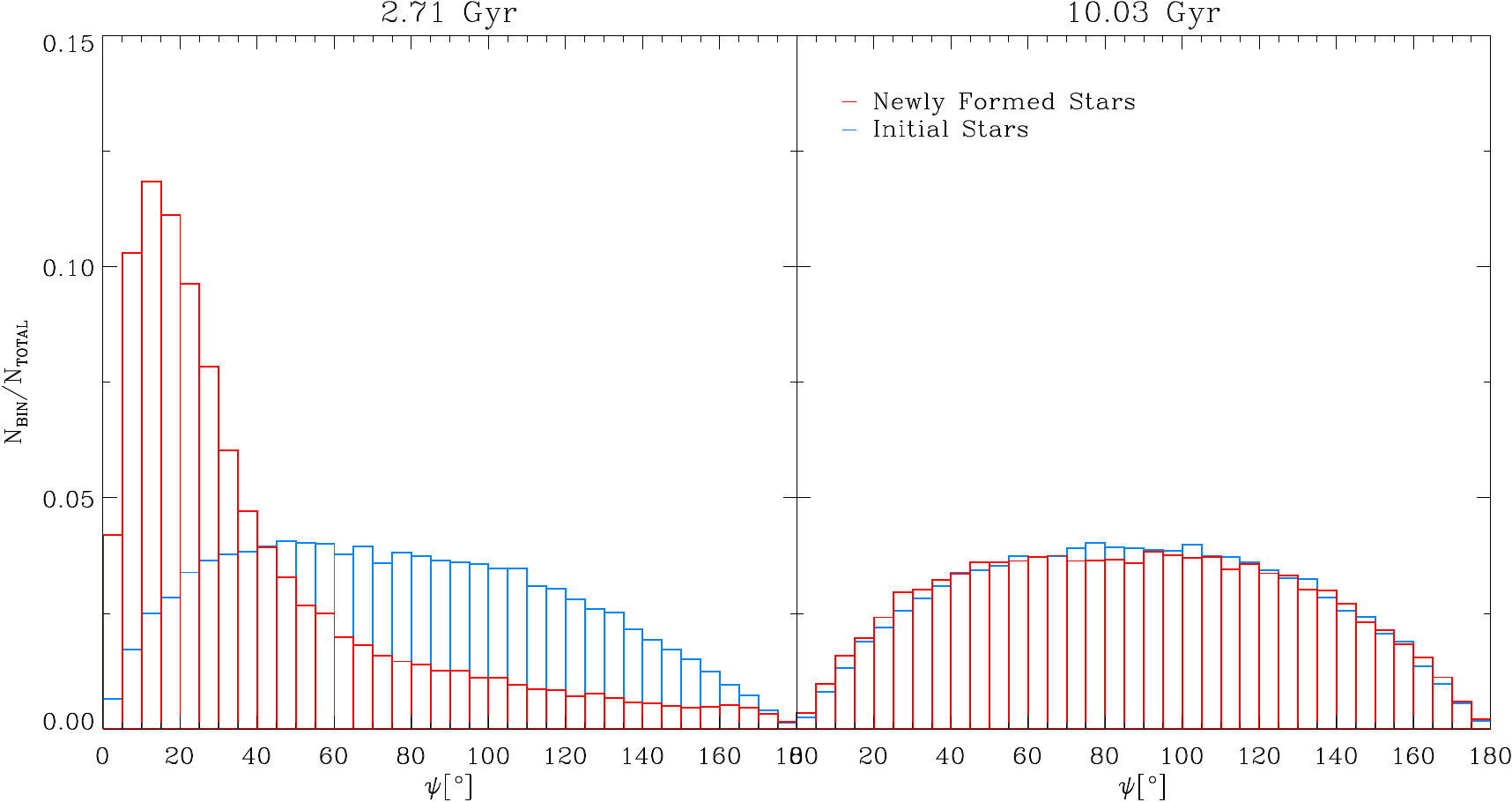}
\caption{Distribution of the misalignment angles between the total stellar angular momentum vector of the KDC after and the angular momentum vectors of the individual stars inside the KDC. The newly formed stars are represented by the red histograms, while the blue histograms illustrate the initial stars.
\textbf{left}: $t=2.71~\mathrm{Gyr}$; \textbf{right}: $t=10~\mathrm{Gyr}$.}
{\label{fig:core_part_angle_histo}}
\end{figure}

To further constrain the fading of the KDC, we take full advantage of the three-dimensional information from the simulation. A useful method to determine
the degree of ordered rotation within a system of particles is to calculate the distribution of the angles between the angular momentum of each particle and
the total angular momentum vector of the complete system. For a rotating system, the distribution of these angles is expected to feature a peak at small angles, while a completely dispersion-dominated system is expected to show a random distribution. The distribution of  {these} angles at two points in time for the PCS is displayed in Figure \ref{fig:core_part_angle_histo} separated for the IS (blue) and NFS (red) populations. The highlighted times are chosen such that, at the earlier point ($t=2.76~\mathrm{Gyr}$), the KDC is fully developed, while, at the later time ($t=10~\mathrm{Gyr}$), the KDC is completely dispersed.

The angle distributions shown in  {Figure} \ref{fig:core_part_angle_histo} conclusively  {confirm} the kinematical difference between the IS and NFS population already found in Section \ref{dissect}.
Already at $t=2.7~\mathrm{Gyr}$, the IS component shows a nearly randomised distribution of angles, indicating a dispersion dominated system.
In~contrast, the distribution for the NFS population completely changes its shape over the period of $7.3~\mathrm{Gyr}$. At~$t=2.7~\mathrm{Gyr}$, the distribution features a clear peak at low angles, indicating ordered rotation in the system. At $t=10~\mathrm{Gyr}$, the alignment disappears entirely, evolving into a purely
random distribution. Therefore, we conclude that the population of NFS generating the KDC signal at earlier times disperses in its further
evolution. This is most likely caused by the interaction between the PCS and the dispersion dominated TCS component. We conclude that, depending on the mass ratio between the two populations, gravitational mixing drives the combined system towards a homogeneous distribution in phase space, dissolving the KDC.
\section{Conclusions}
Results from highly resolved {IFU} observations suggest a dichotomy of KDCs found in the
centres of galaxies. We present an extensive
case study of a KDC formed in a high resolved 1:1 binary disc galaxy merger simulation. We identify the KDC to be consistent with the class of young and
compact KDCs, where the rotational signal is mainly generated by stars that are formed {in situ} during the merger event. Furthermore, the dissection of the KDC reveals a significant dispersion dominated population of stars following highly elliptical orbits permeating the KDC.

We trace the kinematical properties of the KDC over cosmic time, revealing a global gyroscopic precession motion of the KDC.
From this result, we infer a superimposed motion of the KDC consisting of an intrinsic rotation and a global precession of the complete system.
We suspect the precession to be induced by the orbital angular momentum of the merger retained in the gaseous component.

Furthermore, we demonstrate that the KDC is stable for about $3~\mathrm{Gyr}$ after the merger event.
We show that the amount of ordered motion within the KDC subsequently drops significantly on a timescale of
$1.5~\mathrm{Gyr}$, leading to a dispersion of the KDC. This dispersion happens on a similar timescale on which the global precession of the KDC fades.
We suspect that the effect of gravitational mixing between the rotational component and the intruding, permeating dispersion-dominated population causes the KDC to disperse.
Therefore, we conclude that the visibility of a young KDC in the cores of early-type galaxies indicates a (recent) dominant merger event{; however, a more statistical approach is required for a generalised statement {regarding} this matter.}

\vspace{6pt}

\acknowledgments{{Felix Schulze} gratefully thanks the complete Cast Group at the University Observatory Munich for helpful discussions. The authors thank Peter Johansson for running the simulation and providing the output data.
{Felix Schulze} also acknowledges the Max Planck Institute for Extraterrestrial Physics for providing the funding to attend the conference \textit{On the Origin and Evolution of Baryonic Galaxy Halos}.}

\authorcontributions{F.S. and R.-S.R. analysed the data and wrote the paper. K.D. helped to physically interpret the results and provided useful
input during the project.}

\conflictsofinterest{The authors declare no conflict of interest.}

\bibliographystyle{mdpi}
\reftitle{References}

\end{document}